# Quantized conductance of a suspended graphene nanoconstriction

*Nikolaos Tombros,[1,2]\* Alina Veligura,[2] Juliane Junesch,[2] Marcos H. D. Guimarães,[2] Ivan J. Vera Marun,[2] Harry T. Jonkman,[1] and Bart J. van Wees[2]*

[1] Molecular Electronics, Zernike Institute for Advanced Materials, University of Groningen, The Netherlands

[2] Physics of Nanodevices, Zernike Institute for Advanced Materials, University of Groningen, The Netherlands

\*e-mail: n.tombros@rug.nl

A yet unexplored area in graphene electronics[1] is the field of quantum ballistic transport through graphene nanostructures. Recent developments in the preparation of high mobility graphene[2,3,4] are expected to lead to the experimental verification and/or discovery of many new quantum mechanical effects in this field. Examples are effects due to specific graphene edges, such as spin polarization at zigzag edges[5] of a graphene nanoribbon[6,7] and the use of the valley degree of freedom in the field of graphene valleytronics[8]. As a first step in this direction we present the observation of quantized conductance[9,10] at integer multiples of $2e^2/h$ at zero magnetic field and 4.2 K temperature in a high mobility suspended graphene ballistic nanoconstriction. This quantization evolves into the typical quantum Hall effect for graphene at magnetic fields above 60mT. Voltage bias spectroscopy reveals an energy spacing of 8 meV between the first two subbands. A pronounced feature at $0.6 \times 2e^2/h$ present at a magnetic field as low as ~0.2T resembles the "0.7 anomaly" observed in quantum point contacts in a GaAs-AlGaAs two dimensional electron gas, having a possible origin in electron-electron interactions[11].

Conductance quantization in zero magnetic field in graphene ribbons is expected to strongly depend on the type of edge termination[6, 7, 12, 13, 14]. In the case of ideal non-disordered armchair edges the valley degeneracy is lifted, leading to a quantization sequence 0 (for a semiconducting ribbon), 1, 2, 3, … ×$G_0$, when the Fermi energy is raised or lowered from the charge neutrality point. Here $G_0 = 2e^2/h$ with $e$ the electron charge, $h$ the Planck constant and the factor 2 is due to the spin degeneracy. For zigzag edges on the other hand, theory predicts a quantization in odd multiples 1, 3, 5,… ×$G_0$, reflecting the presence of both spin, as well as valley degeneracy. However, realistic devices have a finite (edge) disorder which will dominate the electronic transport in long and narrow ribbons, making the experimental observation of conductance quantization very challenging. Signatures of the formation of one-dimensional subbands due to quantum confinement have been reported for nanoribbons fabricated on a silicon oxide ($SiO_2$) substrate[15, 16]. However those devices are not in the ballistic regime since they have the characteristics of a diffusive, disordered system and lack uniform doping due to strong interaction with the substrate. In such a narrow and long ribbon an edge disorder of typically only a few percent of missing carbon atoms will prevent the observation of quantum ballistic transport and conductance quantization[17, 18, 19].

A way to circumvent this problem is to prepare a constriction with a length comparable or shorter than the width, for which conductance quantization is theoretically possible for an edge disorder of 10% or even higher[18, 19, 20]. In order to investigate quantum ballistic transport and conductance quantization in graphene it is therefore crucial to prepare a narrow, short and high mobility constriction with uniform (gate controlable) doping. This can be achieved by decoupling the graphene layer from the substrate and preparing a high mobility graphene layer suspended 0.2-1 µm above the $SiO_2$ surface. High quality quantum Hall effect (QHE) and fractional QHE were measured experimentally in such devices using a 2-probe geometry.[21, 22]

In this work we prepared similar 2-probe devices using a newly developed polymer based method[23] (see **Methods**) resulting in suspended graphene at 1 µm distance above the SiO$_2$/Si substrate (Fig. 1a, b). The suspended graphene layer is contacted by 80 nm thick titanium/gold electrodes supported by 1µm thick pillars of LOR-A polymer. The electrical characterization was performed using a standard lock-in technique with an applied current of 2.5-10 nA. Application of a voltage to the Si substrate underneath the 500nm thick SiO$_2$ allows us to tune the charge carrier density in the suspended graphene device. In order to obtain high mobility it is crucial to remove the polymer (and possibly other) contaminants present on the suspended graphene after fabrication. For this we anneal the graphene layer by sending a DC electrical current through it (~1mA/µm) in vacuum at 4.2 K [2, 24], which leads to local Joule heating and to an estimated temperature of ~500°C.[2] Mobilities as high as 600.000 cm$^2$/Vs at a charge carrier density of 5 ×10$^9$ cm$^{-2}$ at 77K have been reported in such devices, indicating that the electron mean free path can be several hundred nanometers long[23]. What makes the current annealing step special is that it not only can lead to a high mobility sample, but it can also result in the formation of nanoconstrictions (see Fig. 1a). Note however that although we can systematically obtain high mobility graphene devices with a typical yield of 20% (see **SI**), the formation of these nanoconstrictions is still not well controlled.

We nevertheless succeeded to do electronic measurements of a high mobility graphene nanoconstriction with uniform doping showing conductance quantization at zero magnetic field[9,10] for both electrons and holes. For this device we plot the conductance $G$ at 4.2K for holes versus the Fermi wavenumber $k_F$ in Fig 2a. The Fermi wavenumber $k_F = \sqrt{\pi n}$ is determined by the gate voltage applied to the Si substrate, which allows us to tune the density of charge carriers $n$ in a continuous way from 0 to ~3 ×10$^{11}$cm$^{-2}$. The formation of quantized plateaus at 1, 2 and 3 ×$G_0$ is visible, and also the development of the plateau-like features at 4 and (possibly) 5 ×$G_0$. Note that the presence of quantization at both odd and even multiples of $G_0$ implies that the valley degeneracy is lifted. Although with slightly lower quality, similar plateaus are also observed for electrons (Fig. 2b). The initial width of this device before the current annealing step was about 2.5 µm and it is suspended over 1.5 µm distance between the gold electrodes (see **SI**). An estimate of the actual width $W$ of the constriction formed after the current annealing step can be obtained using the approximate semiclassical relation $G_{bal} \approx 4e^2/h * k_F W/\pi$ for ballistic graphene constrictions. From this relation we extract $W \approx 200$ nm for holes and 275 nm for electrons. This difference in obtained widths is probably related to the uncertainty in the exact position of the Dirac point. This is due to the presence of small non-uniform residual doping, which also results in different confinement potentials for electrons and holes. This can also account for the different quality of the quantized plateaus.

To confirm our conclusions we studied the transition to the QHE by applying a perpendicular magnetic field **B** at 4.2K (Fig. 3a)[25]. The typical QHE behavior for graphene, showing quantized plateaus at 1, 3, 5,… $\times G_0$ is observed when the magnetic field is strong enough that the electron (hole) cyclotron diameter is smaller than the width of the constriction. The situation changes when the field strength is reduced such that the cyclotron diameter $2l_c$ becomes equal or larger than the width of the constriction. In this case the carriers start experiencing (quantum) confinement and a continuous crossover is expected from the QHE regime to quantized conduction at zero magnetic field[25, 26]. The edge channels which carry the current in the quantum Hall regime continuously transform into one dimensional subbands at zero field. This effect is clearly visible in figure Fig. 3b where the $G_0$ plateau remains well developed down to 0mT. The distance $\Delta V_N$ in gate voltage between the center of the quantized plateau corresponding to the first subband (N = 1) and the Dirac neutrality point versus magnetic field is shown in the Fig. 3c. At magnetic fields above 60mT we see the linear scaling of the plateau position in the gate voltage (or density) with magnetic field, characteristic of the QHE regime. However, at B ≈ 60mT we have a crossover below which we observe a saturation in the position of the plateau. Using the relation $l_c = \hbar k_F/eB$ and $2l_c = W$ (which holds at the crossover) at 60mT for N = 1 and 150 mT for N = 3 (Fig. 3d), we extract W ≈ 300 nm. Although one has to be careful to apply these semiclassical relations in the quantum regime, this width is consistent with the width extracted from the fit of $G$ versus $k_F$ in Fig. 2.

Surprisingly, in the magnetic field traces at 0.25, 0.5 and 1T (Fig. 3b) we observe a well developed feature at ≈ 0.6×$G_0$ which strongly resembles the characteristic "0.7 anomaly" observed at ≈ 0.7×$G_0$ in quantum point contacts in GaAs-AlGaAs heterostructures[11]. This feature cannot be the result of Zeeman splitting since at 200 mT this splitting is only $g\mu_B B$ ≈ 25µeV (g ≈ 2) and an order of magnitude smaller than the thermal energy at 4.2 K. We attribute the observed effect to electron-electron interactions similar to the case for the "0.7 anomaly" in GaAs-AlGaAs[11,27]. This unique opportunity to study electron-electron interactions in a graphene nanoconstriction at a moderate field of a few hundred mT or lower is complementary to the high magnetic field studies done recently in graphene[21,22]. At fields above 2T other features develop which could be precursors of the FQHE in a constriction (Fig. 3a).

Finally we perform voltage spectroscopy measurements in order to extract the subband energy spacing. In Fig. 4a, we present the measurement at B = 0T and in Fig. 4b at B = 500mT. The results at B = 500mT show a conductance quantization at 1, 3, 5 and 7 ×$G_0$ at zero or low voltage bias. The formation of a half integer quantized plateau[28] in between the N = 5 and N = 7 plateaus in Fig. 4b is observed for a bias of approximately 8 meV. This is close to the expected value of 7.5 mV corresponding to the average energy spacing between N = 5 and N = 7 plus the average energy spacing between N = 7 and N = 9 for subband energies $E_n = v_F\sqrt{2\hbar eBn}$, (where $n$ = 0, 1, 2... indicates the orbital quantum number). After this control measurement we extract in a similar way the energy spacing (Fig. 4a) at B = 0T. The results show quantization at 1, 2, 3 and 4 ×$G_0$ at zero bias. Here we extract an energy spacing of 8 meV between the N = 1 and N = 2 subbands which corresponds to an energy spacing $\Delta E = \hbar v_F \pi/W$ = 8 meV when we assume a 240 nm wide constriction. These energy scales are consistent with the observed weak temperature dependence of the quantized conductance at 1.5, 4.2 and 12K.

Although we have verified in 3 independent ways the formation of a constriction with a width of about 250 nm, we could not confirm the width after the measurement with e.g. scanning electron microscopy, because the constriction broke during warming up to room temperature. The brittleness of suspended graphene nanoconstrictions is a known problem, which might be solved by preparing high mobility constrictions on a crystalline substrate like boron nitride.[4]

In conclusion, we have shown quantized conductance in a quantum ballistic graphene nanoconstriction. The appearance of quantized conductance at integer multiplies of $G_0 = 2e^2/h$ at zero external magnetic field is assigned to a constriction in which the valley degeneracy is lifted. The quantization found at $\approx 0.6 \times G_0$ at finite magnetic field resembles the $0.7 \times G_0$ anomaly observed in quantum point contacts in a GaAs-AlGaAs 2DEG. Future experiments in graphene nanoconstrictions can shine light on the detailed role of edges in the effective scattering of ballistic charge and spin carriers at zigzag or armchair edges and the effect of strain on quantized conductance[29, 30].

**Methods**

**Sample Preparation**

The preparation of our devices is done as in Ref. 23. We use an acid free method in order to fabricate a suspended graphene device. For this we spin coat a 1.150 μm thick LOR-A (MicroChem) resist layer on a highly n-doped (0.007 Ωcm) 4" Si wafer covered with 500nm silicon oxide dielectric. The highly doped Si wafer is used as an back gate of our suspended graphene device. We deposit HOPG graphene on the LOR-A polymer using the scotch tape technique and use standard electron beam lithography (EBL) in order to contact the graphene layer to metallic electrodes. We evaporate 5 nm of Ti as adhesion layer and 75 nm of Au using an e-gun evaporator a pressure of 5.0 10$^{-7}$ mbar. After lift-off in hot (80$^{\circ}$C) xylene we perform a second EBL step in order to expose the LOR resist underneath the graphene layer. We develop in ethyllactate to remove the EBL exposed LOR resist, rinse the sample in hexane and blow it dry gently with nitrogen. The current annealing technique used to improve the quality of our device is is described in the **SI**.

**Acknowledgments**

We would like to thank Bernard Wolfs and Johan G. Holstein for technical assistance, Caspar van de Wal and Tamalika Banerjee for usefull discussions. This work is part of the research program of the Foundation for Fundamental Research on Matter (FOM) and supported by NanoNed, NWO and the Zernike Institute for Advanced Materials.

**Author contributions:**

N.T., A.V., J.J fabricated the devices, performed the electronic measurements and analyzed the data. I.J.V.M. developed the current annealing method for our devices and M. H. D. G. contributed to the DC bias spectroscopy measurements and analysis of data. N.T, H.T.J. and B.J.W. supervised the experiments and analysis of the results. N. T. wrote the paper with contributions from all authors.

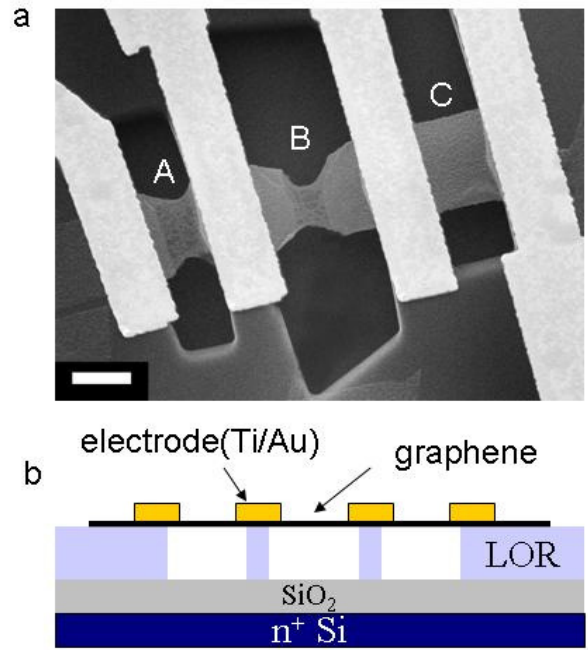

**Figure 1**. **A typical suspended graphene device.** a) Scanning electron microscopy picture of a typical suspended high mobility graphene device showing the formation of graphene constrictions after the current annealing step in vacuum at 4.2K (regions A and B). The scale bar is 2 µm. No current annealing was applied to region C. b) A schematic cross-section of the device. The graphene layer is suspended about 1 µm above the 500 nm thick $SiO_2$ and the electrodes are kept in place by pillars of LOR polymer. The n+ doped silicon substrate is used as a back gate electrode in order to control the charge-carrier density.

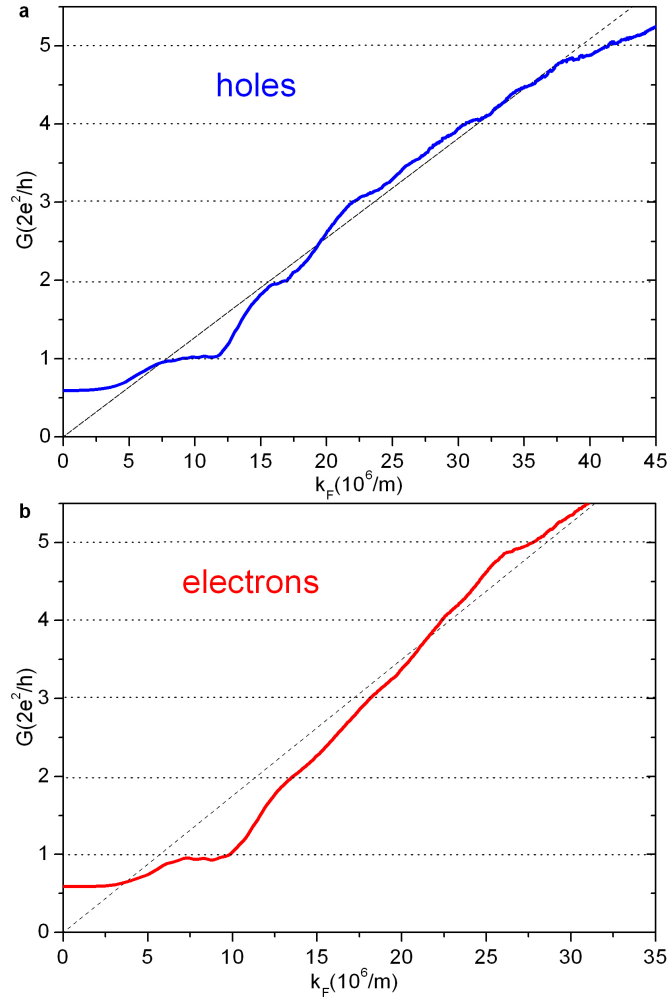

**Figure 2**. **Quantized conductance in a graphene nanoconstriction at T = 4.2K at zero external magnetic field** a) Conductance *G* as function of the Fermi wavenumber $k_F$ at zero external magnetic field for holes. A total of 80Ω contact resistance was subtracted. Clear quantization is observed at 1, 2, 3 x $2e^2/h$ and plateau like features are visible at 4 and 5 x$2e^2/h$. The dashed line is a fit using the semiclassical relation G = $4e^2/h$*$k_F$W/π in the ballistic regime which gives the width of the constriction $W \approx 200$ nm b) For electrons we observe a similar sequence of conductance quantization (80Ω contact resistance was subtracted). In this case we obtain $W \approx 275$ nm from the fit.

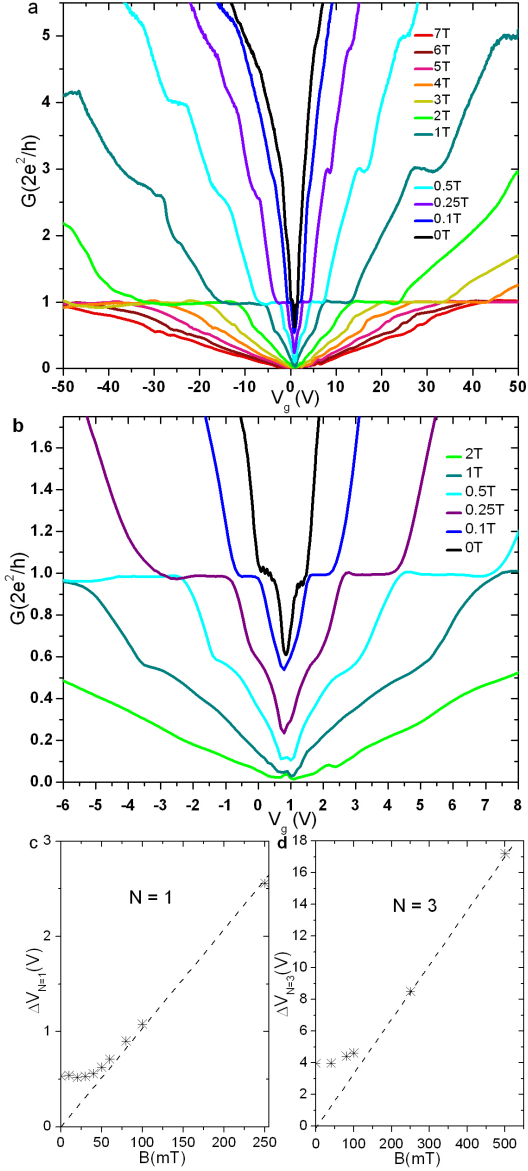

**Figure 3**. **Quantized conductance in a graphene nanoconstriction in zero and finite magnetic field at T = 4.2K** a) Magnetic field dependence (in Tesla) of the two probe quantum Hall effect in a graphene ballistic nanoconstriction as a function of the gate voltage $V_g$. The capacitance of 8 aF/μm$^2$ is extracted from the 2-probe quantum Hall measurements at a magnetic field of 500mT. The quantized plateaus at 1, 3 and 5 x2e$^2$/h are characteristic for graphene. A contact resistance of 80Ω was subtracted from the 2-probe measurements. b) Transition from quantum Hall effect to quantized conductance at zero magnetic field. Note that at zero magnetic field we observe Fabry-Pérot like oscillations superimposed on the 2e$^2$/h plateau, possibly the result of reflection at the ends of the constriction. The feature at 0.6 x2e$^2$/h
1313

(e.g at Vg = -1.5V and 3V for B = 0.5T) is believed to be the result of electron-electron interactions, similar to the "0.7 anomaly" observed in a GaAs-AlGaAs heterostructures (see main text) c) The distance $\Delta V_N$ in gate voltage between the center of quantized plateau corresponding to the N = 1 subband and the charge neutrality point versus external magnetic field ***B*** applied perpendicular to the suspended graphene layer. Here, $\Delta V_N$ saturates at 0.5V for fields below 60mT from which we extract the width of the constriction (300nm). d) The same plot was made for the N = 3 subband and approximately the same width was obtained. Note that $\Delta V_N$ saturates below 150mT in this case.

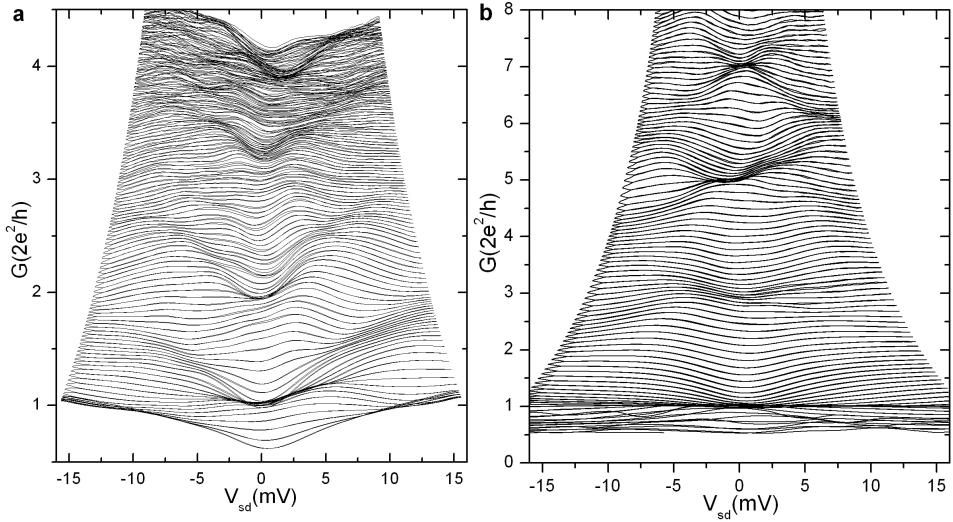

**Figure 4**. **Voltage bias spectroscopy at T = 4.2K.** a) The differential conductance $G$ versus DC bias voltage $V_{sd}$ measured with an excitation AC voltage of $V_{ac} = 150\mu V$ in the gate voltage interval of -6V $< V_g <$ 0.8V. Each line in this plot corresponds to a DC bias measurement at a different gate voltage, from $V_g$ = -6V (top) to 0.8V (bottom) in steps of 50mV. At zero external magnetic field and $V_{sd}$ = 0V we observe conductance quantization at 1, 2, 3 and 4 x $2e^2/h$. The energy spacing between the N = 1 and N = 2 subbands is approximately 8 meV which is consistent with the energy spacing expected for a 240 nm wide constriction. b) Voltage spectroscopy at B = 500mT and -40V $< V_g <$ 0.8V. The regular 1, 3, 5 and 7 $2e^2/h$ plateaus are obtained (after subtraction of 700Ω). The energy spacing between the N = 5 and N = 7 subbands is approximately 8 meV (see main text).

# Quantized conductance of a suspended graphene nanoconstriction

## Supplementary Information

### 1. Current annealing

Current annealing of suspended graphene membranes was performed by ramping up the DC current across the suspended graphene devices for each measured region separately in vacuum (2.0 10$^{-7}$ mbar) at a temperature of 4.2K. While increasing the DC current through the devices the resistance of the device was monitored. At typical current densities of approximately 7 A/cm (about a current of 1.5 mA) the resistance starts increasing rapidly, indicating the combination of two effects: the increase in the graphene temperature (to T > 500$^o$C), followed by the shift of the charge neutrality point from a highly doped state (usually p-doped) towards zero gate voltage. The current density required to clean the graphene membranes varies from sample to sample and depends on the length and width of the graphene. We relate this to the fact that suspended graphene cools down via the metal contacts, the closer they are the higher the current density required to bring the charge neutrality point to zero (hence reach high enough temperature for desorbing polymer remains from the graphene surface). In Fig.S1 a scanning electron microscopy (SEM) image of a typical suspended device is shown. Regions A and B are annealed with current densities of 6.8 and 4.8 A/cm respectively. For comparison the region C was left untouched and shows highly p-doped state. In the SEM picture one can see the difference between the current annealed and non-annealed regions. The region C shows a homogeneously coverage of residues, which is not observed in the annealed regions. Note also that the graphene layer has a tendency to constrict after current annealing as in region A (see also Fig.1a in the main text). In several cases the devices break during the annealing procedure. About 20% of the 2-probe regions survive the current annealing step and become high mobility samples.

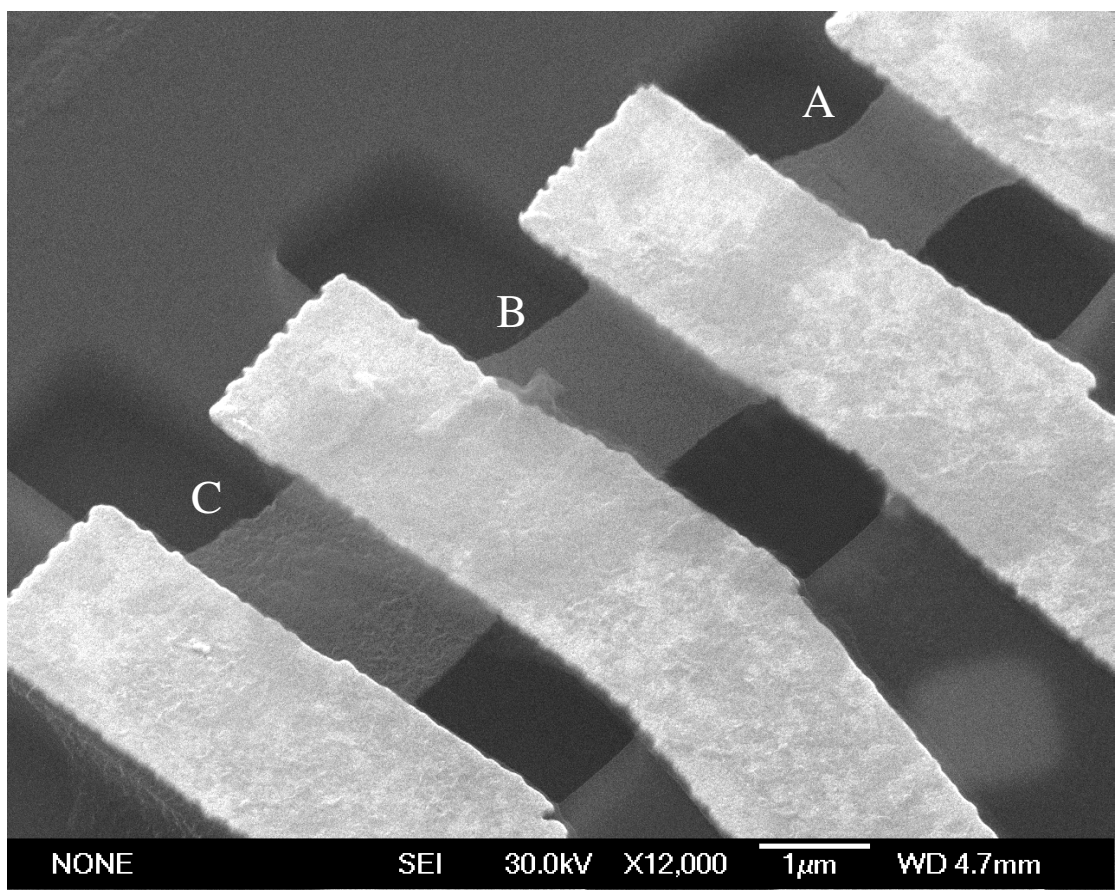

**Figure S1.** SEM image of a typical suspended device on LOR polymer. Regions A and B were annealed with DC current while C was left untouched for comparison.

## 2. Overview of measured devices

We fabricated about 20 two-probe devices with different dimensions. Here we present the results of electronic transport close to the ballistic regime for 4 different samples. In Figures S2, S3, S4 and S5 we present the resistance and conductance as a function of gate voltage $V_g$ (a and b respectively), c) the conductance as a function of the Fermi wavenumber $k_F$ and d) the mean free path ($\lambda$) of the charge carrier versus $k_F$ for each sample (#1 to #4)

For each separate case the capacitance of the system ($\alpha$) was determined from the filling factors in the quantum Hall regime. In the calculations we used $n = \alpha\ (V_g - V_D)$, where n is the induced charge carrier density and $V_D$ is the position of the charge neutrality point. The Fermi wavenumber was obtained from the relation $k_F = \sqrt{\pi n}$. We extract the mean free path of the charge carriers using the Einstein relation for conductivity $\sigma = \nu\ e^2 D$, where $\nu$ is the density of states for a single graphene layer and the diffusion constant in two-dimensions is given by $D = \frac{1}{2} v_F \lambda$, with $v_F$ is Fermi velocity and $\lambda$ the mean free path. Substituting $\nu = \frac{g_v g_s 2\pi |\varepsilon|}{h^2 v_F^2}$ and $|\varepsilon| = \hbar v_F k_F$ we obtain $\lambda = \frac{\sigma h}{2 e^2 k_F}$.

The length L and width W of the samples are indicated in each of the figures S2, S3, S4 and S5 in panel a). Since most of the measurements were performed in 2-probe geometry we subtracted the contact resistance (typically 50Ω per contact). For each sample we calculate the number of expected one-dimensional modes (N) through the channel $N = \frac{k_F W}{\pi}$ at a specific value of $k_F$ and also the corresponding ballistic conductance $G_{bal} = \frac{4e^2}{h} \frac{k_F W}{\pi}$

#1

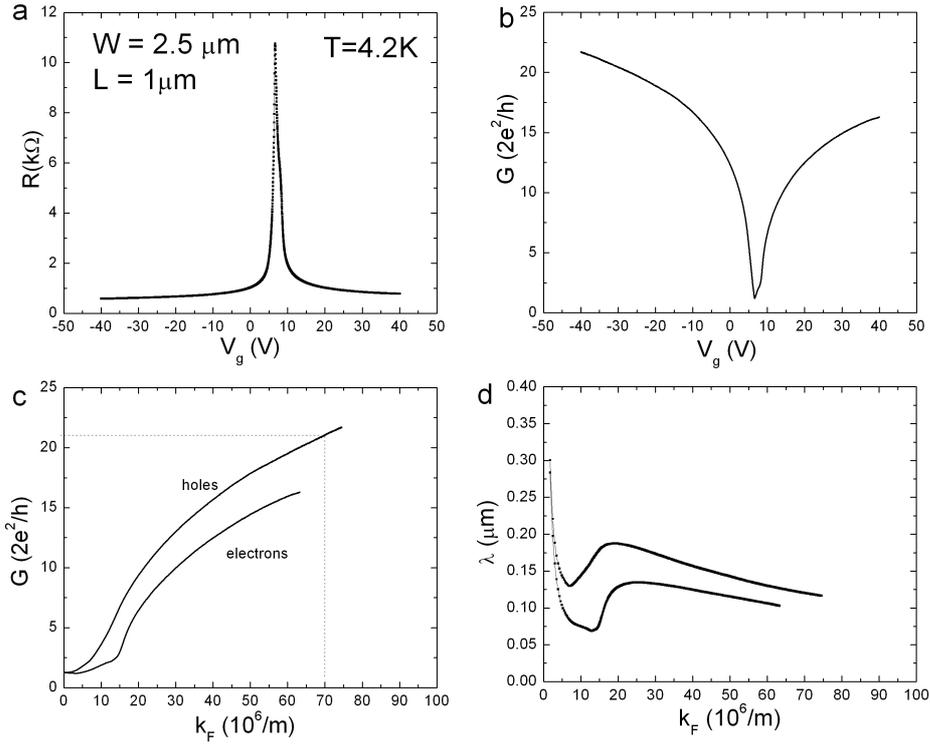

**Figure S2.**

For sample #1 (Fig S2) the calculated value of conductance at $k_F = 70\times10^6$/m according to its width (2.5 µm) is 110 $\times 2e^2/h$, while the measured value was ~21 $\times 2e^2/h$. The transmission of the channel is around 19% which corresponds to a mean free path of 190 nm in agreement with the mean free path extracted from the Einstein relation (Fig. S2 d).

#2

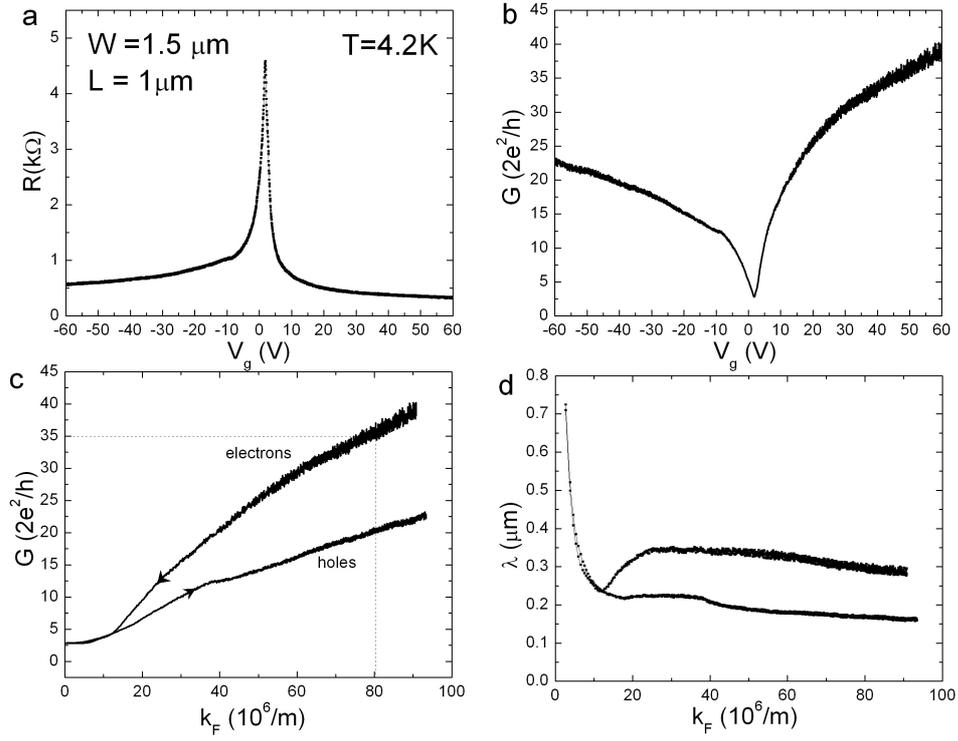

**Figure S3.**

The same calculations of the conductance for Sample #2 (Fig S3) at $k_F = 80\times10^6$/m and for the width (1.5 µm) results in $G = 75 \times 2e^2/h$, while the measured value was $G \sim 35 \times 2e^2/h$, which means that the transmission of the channel is about 45% and a mean free path of 450 nm.

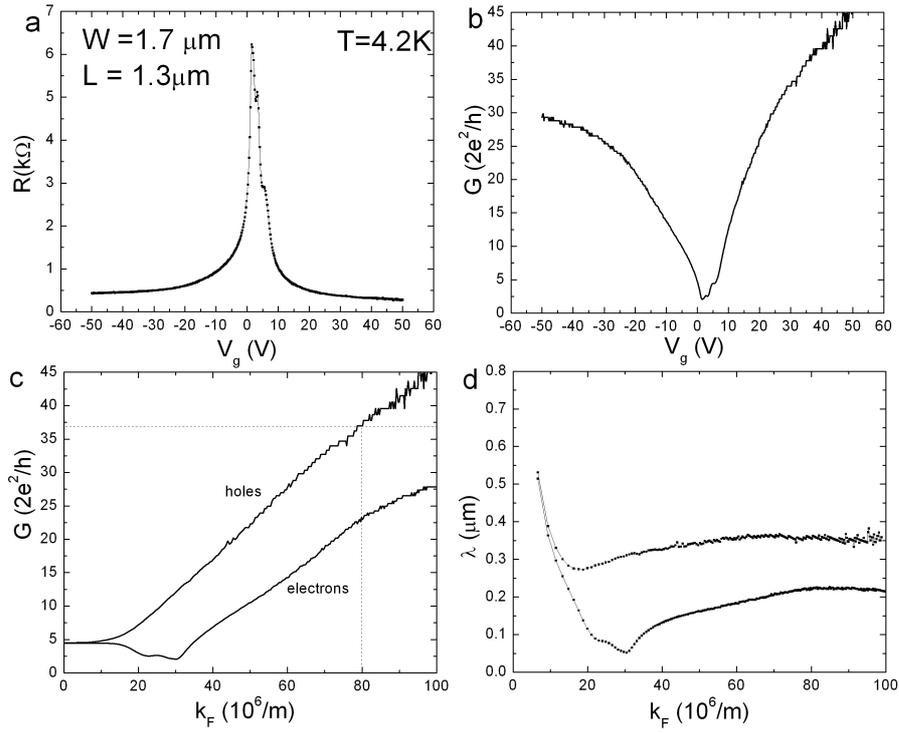

**Figure S4.**

For Sample #3 (Fig S4) at $k_F = 80 \times 10^6$/m and for the width (1.5 µm) $G = 85 \times 2e^2/h$, while the measured value was $G \sim 37 \times 2e^2/h$, giving a transmission of 43% and a mean free path of 430 nm.

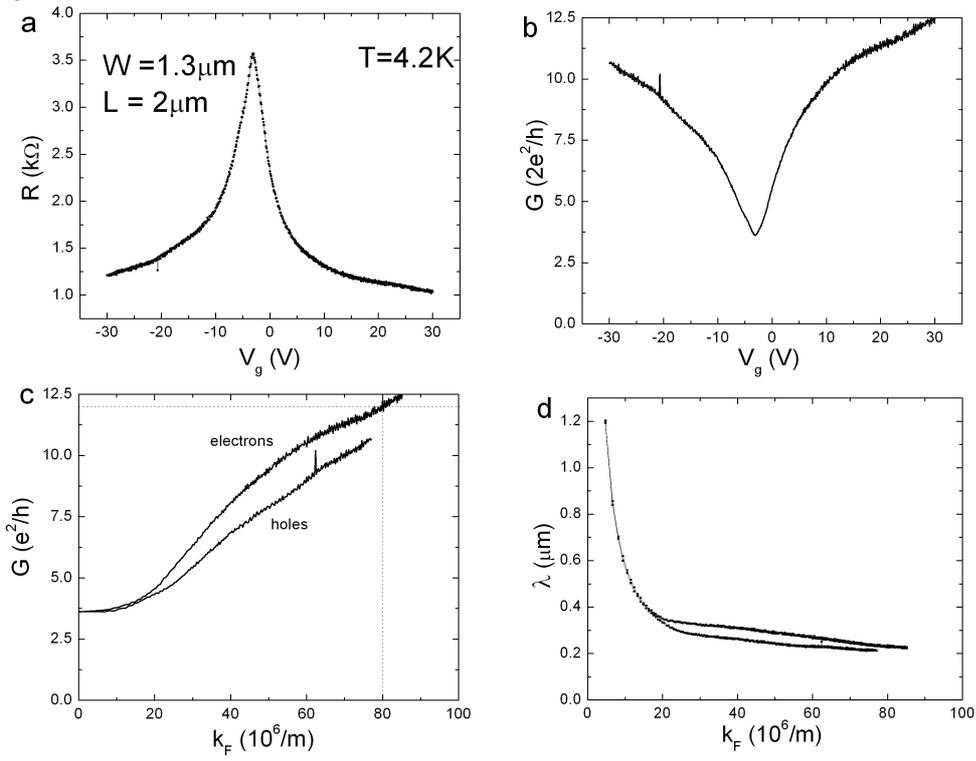

**Figure S5.** Sample #4

According to the number of calculated modes, sample #4 (Fig. S5) has 18% transmission and a mean free path of 360 nm at $k_F = 80\times10^6$/m

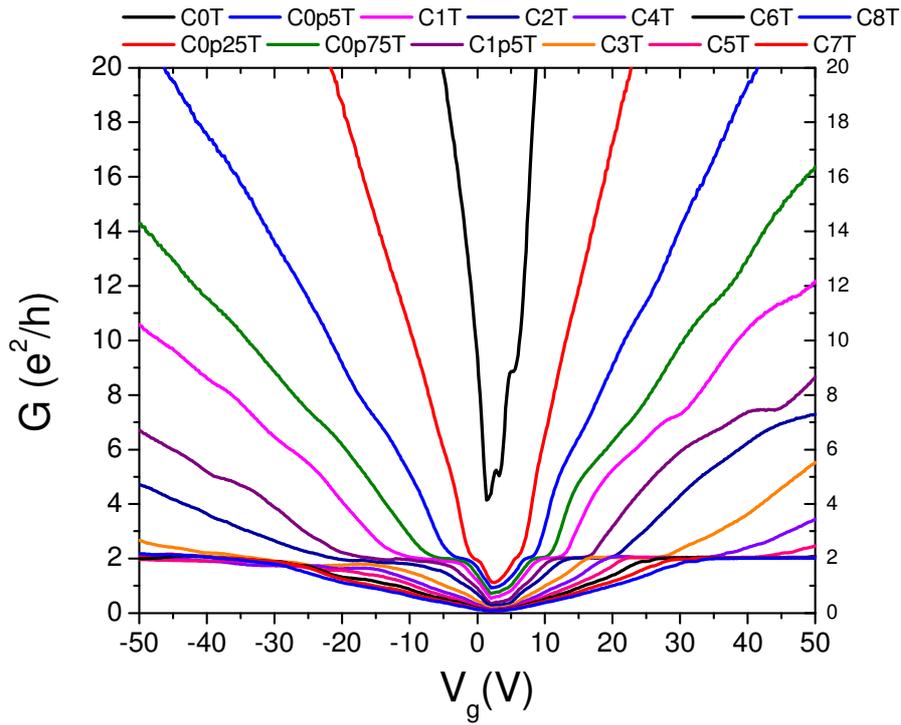

**Figure S6.** Observed QHE in sample #3.

In Figure S6 we show the QHE for Sample #3. The $2e^2/h$ plateau is observed down to a magnetic field of 250 mT indicating high quality graphene. However, the position of the Dirac point at 2.5V and the non-well developed plateaus at 6, 10 and 14 $e^2/h$ indicate some inhomogeneity in the residual doping. Note that in this sample the $2e^2/h$ plateau does not extend down to 0T in contrast to the sample discussed in the main text.

Table 1 Dimensions and maximum resistance $R_{max}$ for different samples

| Sample name | Dimensions, µm | $R_{max}$ (kΩ) after annealing |
|---|---|---|
| Sample showing quantized conductance for both electrons and holes (see main manuscript) | L=1<br><br>W=2.5 before annealing<br><br>W= 0.3 after annealing | 21.7 |
| Sample #1 | W=2.5; L=1 | 10.5 |
| Sample #2 | W=1.5; L=1 | 4.5 |
| Sample #3 | W=1.7; L=1.3 | 6.2 |
| Sample #4 | W=1.3; L=2 | 3.5 |

In table 1 we show the resistance $R_{max}$ at the charge neutrality point after the annealing step. The highest resistance was obtained for the sample presented in the manuscript which showed quantized conductance as a result of formation of a constriction. We note that the position of the Dirac point for the device presented in the manuscript is found at 0.8V. This is much closer to 0V as compared to the devices #1-#4 and this points to the fact that there is a very small inhomogeneity due to residual doping.

In conclusion, the devices fabricated with the current annealing step show a mean free path of several hundred nanometers at high charge carrier density and even longer at lower density. The device described in the main text has the best transport properties of all investigated devices.

## 3. Magnetic field offset

We can exclude that the conductance quantization we measured at zero magnetic field B is a result of any remaining magnetization of the superconducting magnet for the following reasons:

1) The conductance quantization does not disappear when we scan the range -100mT to 100mT. From the symmetry between +B and –B we conclude that any remaining magnetization from the superconducting magnet (or other sources) is less than 1mT.
2) The conductance quantization at 4.2K persists when the sample is positioned 20 cm above the superconducting magnet (set to zero field).
3) At a temperature of approximately 12 K, at which the magnet is not in the superconducting state anymore, we still measure quantized conductance.